# Microfiber-based few-layer black phosphorus saturable absorber for ultra-fast fiber laser

Zhi-Chao Luo,[1,3] Meng Liu,[1] Zhi-Nan Guo,[2] Xiao-Fang Jiang,[4] Ai-Ping Luo,[1,3] Chu-Jun Zhao,[2] Xue-Feng Yu,[5] Wen-Cheng Xu,[1,3] and Han Zhang[2,*]

[1]*Guangdong Provincial Key Laboratory of Nanophotonic Functional Materials and Devices, School of Information and Optoelectronic Science and Engineering, South China Normal University, Guangzhou, Guangdong 510006, China*

[2]*SZU-NUS Collaborative Innovation Centre for Optoelectronic Science & Technology, and Key Laboratory of Optoelectronic Devices and Systems of Ministry of Education and Guangdong Province, Shenzhen University, Shenzhen, China*

[3]*Specially Functional Fiber Engineering Technology Research Center of Guangdong Higher Education Institutes, South China Normal University, Guangzhou, Guangdong 510006, China*

[4]*State Key Laboratory of Luminescent Materials and Devices (SKLLMD), South China University of Technology, Guangzhou 510640, China*

[5]*Institute of Biomedicine and Biotechnology, Shenzhen Institute of Advanced Technology, Chinese Academy of Sciences, Shenzhen 518055, China*

*\*hzhang@szu.edu.cn*

## Abstract

Few-layer black phosphorus (BP), as the most alluring graphene analogue owing to its similar structure as graphene and thickness dependent direct band-gap, has now triggered a new wave of research on two-dimensional (2D) materials based photonics and optoelectronics. However, a major obstacle of practical applications for few-layer BPs comes from their instabilities of laser-induced optical damage. Herein, we demonstrate that, few-layer BPs, fabricated through the liquid exfoliation approach, can be developed as a new and practical saturable absorber (SA) by depositing few-layer BPs with microfiber. The saturable absorption property of few-layer BPs had been verified through an open-aperture z-scan measurement at the telecommunication band and the microfiber-based BP device had been found to show a saturable average power of ~4.5 mW and a modulation depth of 10.9%, which is further confirmed through a balanced twin detection measurement. By further integrating this optical SA device into an erbium-doped fiber laser, it was found that it can deliver the mode-locked pulse with duration down to 940 fs with central wavelength tunable from 1532 nm to 1570 nm. The prevention of BP from oxidation through the "lateral interaction scheme" owing to this microfiber-based few-layer BP SA device might partially mitigate the optical damage problem of BP. Our results not only demonstrate that black phosphorus might be another promising SA material for ultrafast photonics, but also provide a practical solution to solve the optical damage problem of black phosphorus by assembling with waveguide structures such as microfiber.

## Introduction

Remarkable progress in two-dimensional (2D) nanomaterials in recent years has highlighted the large potentials of 2D materials for the photonic and optoelectronic applications [1-4]. Owing to its wideband absorption, ultrafast carrier dynamics and 2D planar advantage, graphene has attracted considerable attention and been demonstrated in a number of applications, ranging from optical saturable absorbers (SAs) for ultrafast lasers [5-7] to optical modulator [8,9], photo-detectors [10], and optical sensors [11] for future high speed and broadband optical technologies. However, the relative weak absorption in graphene (only 2.3 % of incident light per single layer) leads to very low level of optical modulation depth, which therefore introduces limitations for nonlinear optics where strong light-matter interaction exists. Transition mental dichalcogenides (TMDs), characterized by atomically thin semiconductors of $MX_2$ where M represents for a transition metal atom (Mo, W, etc.) and X for a chalcogen atom (S, Se, etc.), possess higher resonant absorption at specific wavelength, but the optical response mainly occurs in the visible range due to a relatively large band gap [12-16]. Consequently, there remains a gap between semi-metallic graphene and wideband semiconducting TMDs. It is important to note that photonic devices operating at optical communication band require optical materials with band-gap in the range of 1 eV, which does not fit for the band-gap of the conventional well-known 2D materials such as graphene or TMDs. Fortunately, black phosphorus (BP), has recently joined in the family of 2D materials with layer dependent direct band gap from 0.3 eV (bulk) to 1.5 eV (monolayer) [17, 18]. The basic structure of BP is similar to bulk graphite, in which individual atomic layers stacked together by van der Waals interaction [19]. Inside the single layer, each phosphorus atom is covalently bonded with three adjacent phosphorus atoms in order to form a unique puckered honeycomb structure. In bulk form, BP has a band-gap of 0.3 eV due to the interlayer interaction while its band-gap increases with the decrease of the thickness [20]. The emergence of few-layer BPs can bridge the gap between graphene and TMDs for infrared photonics and optoelectronics, particularly for the optical communication devices.

Different from TMDs that have a crossover from the indirect band-gap to the direct band-gap [12], BP always has a direct type of energy band structure regardless of thicknesses, which can be regarded as a significant benefit for ultra-fast photonics and high frequency optoelectronics. Previous work had already verified that BPs could exhibit the behavior of optical bleaching if under strong light illumination, also known as saturable absorption [21, 22]. Short-pulse lasers, including pulses of the order of several microseconds (e.g. Q-switching) and nanoseconds or even few-femtoseconds in durations (e.g. mode-locking) can be initiated and stabilized by the inclusion of a SA to act as a passive optical switch [23-27]. Therefore, few-layer BPs can be potentially developed as a broadband SA operating at long wavelength side.

In essence, BP represents a type of 2D layered crystal with moderate direct band gap that is ideal for broadband optoelectronic devices at optical communication band. However, the lack of stability in air and ease of oxidation prevent BPs towards its future exploration at high power regime [28,29]. A double layer capping of $Al_2O_3$ and hydrophobic fluoropolymer has afforded BP devices with acceptable air-stability, which might overcome the critical material challenge for applications [30]. From the viewpoint of practical applications, it becomes very urgent to develop a solution to tackle the optical damage or air-instablity problem of BPs.

Herein, we employed the "lateral interaction scheme" in order to increase the optical damage threshold of few-layer BPs, and therefore prevent it from oxidation. Unlike perpendicular illumination where light directly interacts with the surface of BP flakes with a laser beam size down to micro-meter and it therefore introduces extra heat, it was found that the lateral interaction, where the evanescent field of light propagates along the surface of few-layer BPs, can allow the device to endure an incident power up to 56 mW in our experiment. Our nonlinear optics measurement based on the balanced twin-detector method shows that this device exhibits saturable absorption response with a saturable average power of ~4.5 mW and a modulation depth of 10.9%. The center wavelength, spectral width, repetition rate, and estimated pulse duration of the mode-locked fiber laser by microfiber-based BP SA are 1566.5 nm, 3.39 nm, 4.96 MHz, and 940 fs at a pump power of ~70 mW, respectively. The ease of handling of such a microfiber-based BP device offers new opportunities for wide photonic applications.

## Synthesis and characterization of few-layer BPs

The liquid phase exfoliation (LPE) is regarded as a simple and effective technique to prepare 2D nanomaterials from the layered bulk crystals towards the few-layer structures [31]. In this work, the few-layer BPs was also prepared through the LPE method. First, bulk black phosphorous (30 mg) was added to N-Methyl pyrrolidone (NMP) solution (30 ml). Then, the mixture was bath sonicated at 40 kHz frequency and 300 W for 10 hours to carry on the liquid exfoliation of bulk BP. After the sonication step, the as-prepared few-layer BPs was centrifuged at 1500 rpm for 10 min and then the supernatant liquor was taken for the testing samples. The as-prepared few-layer BPs NMP solution was shown in Fig. 1(a). In order to show the morphologies of the as-prepared few-layer BPs, the transmission electron microscope (TEM) image was provided. As can be seen in Fig. 1(b), the prepared BPs could be clearly identified to be layered structure with a size of several hundred nanometers. To better characterize the prepared few-layer BPs, Raman spectrum measurement was carried out by a Horiba Jobin-Yvon LabRam HR VIS high-resolution confocal Raman microscope equipped with a 633 nm laser. The measured result is presented in Fig. 1(c). Three Raman peaks corresponding to one out-of-plane vibration mode $A_g^1$ and two in-plane vibration modes $B_{2g}$ and $A_g^2$ are located at 360.8, 437.2 and 465.3 cm$^{-1}$, respectively. As we know, the Raman spectrum is related to the thickness of the BPs, where the $A_g^1$ and $A_g^2$ modes will shift toward each other with the increased thickness [32]. Compared with the standard Raman spectrum of bulk BPs, our measured result suggests that the BPs have been exfoliated down to be several layers.

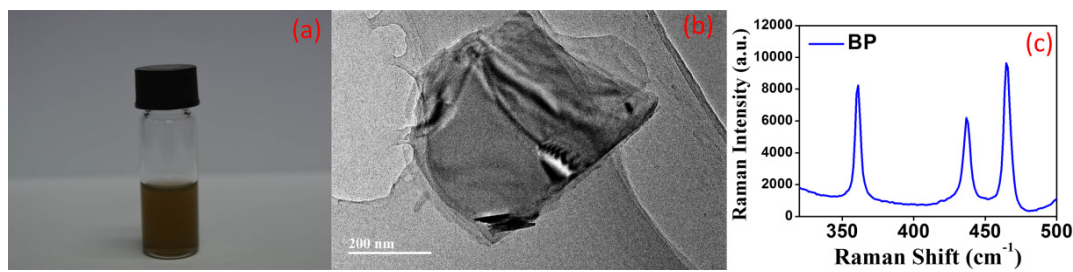

Fig. 1. (a) As-prepared few-layer BPs solution; (b) TEM images; (c) Corresponding Raman spectrum.

To precisely identify the concrete thickness and size of the as-prepared few-layer BPs, the sample were measured through the atomic force microscopy (AFM). The sample was prepared by dropping the BP solution onto the quartz substrate and drying in vacuum. Here, three sections were selected for measuring the thickness of the few-layer BPs, as indicated in Fig. 2 (a). Figures 2(b) to 2(d) summarize the height difference between the substrate and the corresponding three sections. As can be seen here, the absolute thicknesses of all the three sections are at the range of ~0.6 to ~2 nanometers. Considering that the thickness of the single layer BP is ~0.6 nm, it is estimated that the thickness of the as-prepared BPs ranges from 1 to 3 layers. According to the previous findings [33], the band gap of 3-layer BP is about 0.8 eV, which is particularly suitable for fabrication of optoelectronic devices at 1550 nm optical communication band.

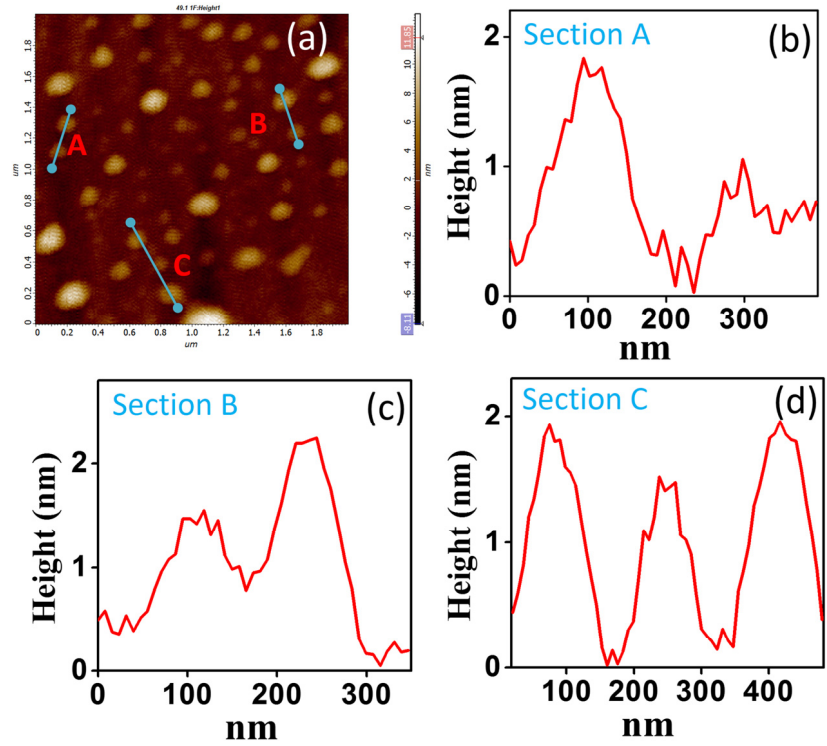

Fig. 2. (a) AFM image of few-layer BPs; (b)-(d) Height profiles of the three sections marked in (a).

## Results and discussions

In order to characterize the nonlinear optical response of the as-prepared BP nanosheets, the Z-scan technique was employed by using a typically experimental setup similar to our previous one [34]. However, the pump source used in this work is a femtosecond laser with 120 fs pulse duration of at a center wavelength of 1550 nm. The laser pulses were generated from optical parametric amplifier (TOPAS-Prime) pumped by a mode-locked Ti:sapphire oscillator seeded regenerative amplifier with a pulse energy of 1.3 mJ at 800 nm and a repetition rate of 1 kHz (SpectraPhysics Spitfire Ace). The average power of the femtosecond pulse was then attenuated to 130 µW for Z-scan experiment. After scanning process of the sample, a normalized Z-scan curve could be obtained through dividing the output beam power by the reference beam power, as shown in Fig. 3(a). From Fig. 3(a), we can see that a sharp peak is located at the beam focus point, which clearly demonstrates that the BP nanosheets possess saturable absorption effect at 1550 nm waveband. Based on the Z-scan curve, the nonlinear saturable absorption curve could be also extracted, as shown in Fig. 3(b). Here, the saturation intensity and the modulation depth are identified to be 2.6 MW/cm$^2$ and ~8.5% respectively.

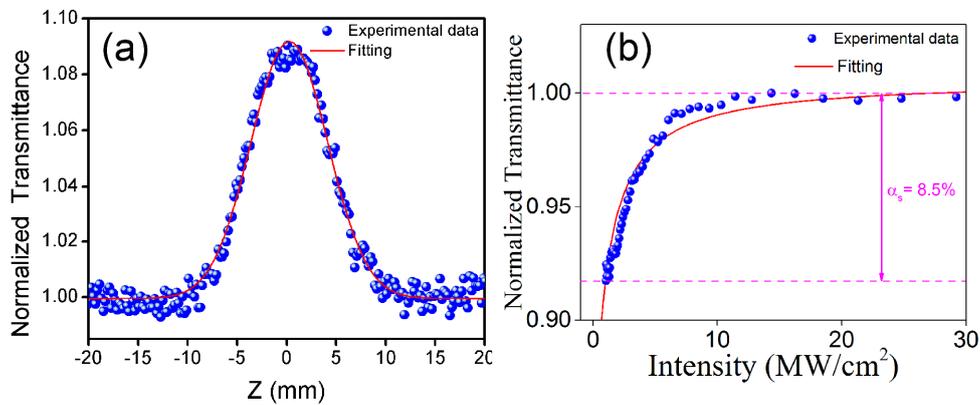

Fig. 3. Nonlinear optical response of the as-prepared few-layer black phosphorus nanosheets. (a) Typical Z scan curve at 1550 nm; (b) The corresponding nonlinear saturable absorption curve.

In the following, we employed the few-layer BPs as SA to construct an ultrafast fiber laser operating at 1550 nm waveband. Firstly, the few-layer BP SA was fabricated. In order to increase the optical damage threshold of few-layer BPs and provide stronger light-matter interaction, we employ an evanescent field interaction scheme of the propagating light with few-layer black phosphorus deposited onto a microfiber. The experimental setup of optical deposition for fabrication of microfiber-based BP SA and the detailed process is similar to that of Ref. [35]. Briefly, the microfiber was drawn from the standard single mode fiber (SMF) with a diameter of ~10 µm. The BP NMP solution with a concentration of 0.08 mg/ml was dripped onto the microfiber and covered the waist region of the microfiber for optical deposition. After the deposition process, the as-prepared microfiber-based BP SA was presented in Fig. 4. As can be seen here, the BP nanosheets (black portion) were well deposited around the microfiber. The deposition length of few-layer BPs is about ~92 μm.

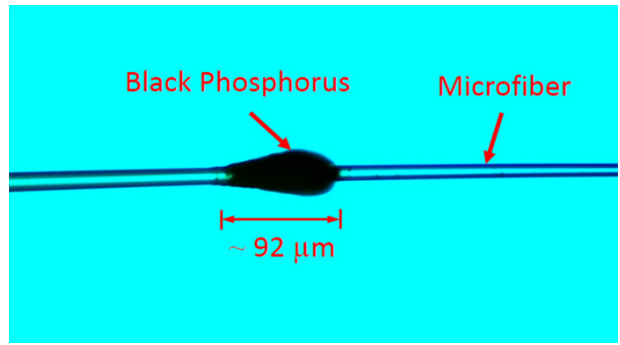

Fig. 4. Microscopic image of the fabricated microfiber-based BP SA.

To check whether the BPs-deposited microfiber still possesses the saturable absorption effect or not, we employed the nonlinear optical response measurement based on the twin-detector scheme. The details of the setup have also been described in Ref. [35]. Here, the pumping femtosecond laser source is an in-house-made femtosecond pulse source (center wavelength, 1554.4 nm; repetition rate, 26 MHz; pulse duration, ~500 fs). By continuously adjusting the input power, the measured optical response was shown in Fig. 5. As can be seen here, the saturable average power and the normalized modulation depth of the BPs-deposited microfiber SA are ~4.5 mW and 10.9%, respectively. Note that the experimentally measured nonsaturable loss is about 54.2%. Therefore, the proposed few-layer BPs-deposited microfiber photonic device could act as a high-performance SA for ultrafast fiber laser to generate ultrashort pulse.

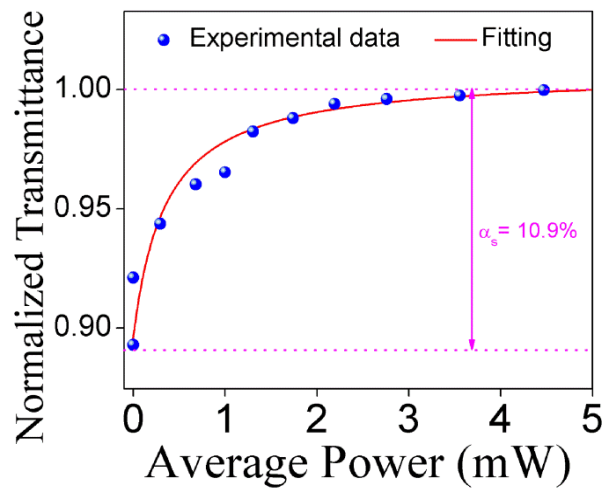

Fig. 5. Nonlinear saturable absorption and corresponding fitting curve of fabricated microfiber-based BP SA.

The schematic of the proposed passively mode-locked fiber laser incorporating microfiber-based BPSA was presented in Fig. 6. A segment of ~7 m erbium-doped fiber pumped by a 980 nm laser diode was used as the gain medium. All the other fibers are standard single-mode fiber (SMF) with a length 34.4 m. Therefore, the fundamental repetition rate is 4.96 MHz. The unidirectional operation of the fiber laser was ensured by a polarization-independent isolator (PI-ISO). The polarization states of propagation light are adjusted through a pair of polarization controllers (PCs). The laser output was taken by a 10% fiber coupler, which is monitored by an optical spectrum analyzer (Yokogawa AQ6317C) and an oscilloscope (Tektronix DSA 70804) with a high-speed photodetector (New Focus P818-BB-35F, 12.5GHz). Moreover, the pulse profiles were measured by a commercial autocorrelator (FR-103XL).

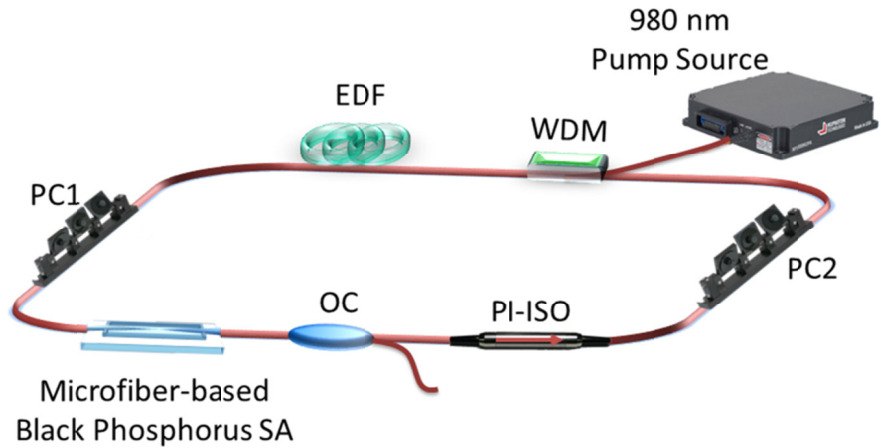

Fig. 6. Schematic of the microfiber-based BP SA mode-locked fiber laser.

By virtue of the saturable absorption effect of few-layer BPs-deposited microfiber device, the fiber laser could achieve the self-started mode-locking state at the pump power of 30 mW. For better performance of the fiber laser, the pump power was then increased to ~70 mW. Figure 5 summarizes the passive mode-locking performance of the few-layer BPs based ultrafast fiber laser. The spectrum of the mode-locked pulse centered at 1566.5 nm with a 3-dB bandwidth of 3.39 nm, as shown in Fig. 7(a). The symmetric Kelly sidebands were clearly presented on the mode-locked spectrum, indicating that the fiber laser operates in the soliton regime with anomalous dispersion [36]. The corresponding mode-locked pulse-train was depicted in Fig. 7(b). The repetition rate is a fundamental one of 4.96 MHz, which was determined by the cavity length. A larger scanning range of pulse-train under 500 μs is also presented in the inset of Fig. 7(b), which demonstrates that no evident power fluctuation could be observed. To identify the duration of mode-locked pulse, we measured the autocorrelation trace correspondingly, as shown in Fig. 7(c). The pulse width was measured to be 940 fs if the Sech$^2$ intensity profile was assumed. Therefore, the time-bandwidth product is 0.38, indicating that the pulse is slightly chirped. In addition, the radio frequency (RF) spectrum of the mode-locked pulses was also measured, as presented in Fig. 7(d). The RF spectrum shows a signal-to-noise ratio of ~50 dB, suggesting that the fiber laser is operating in a stable regime. In order to further investigate the stability of the mode-locked fiber laser, the optical spectrum was recorded every 1-hour, as shown in Fig. 8. Here, it should be noted that the central wavelength and mode-locked spectral bandwidth are reasonably stable over the time period.

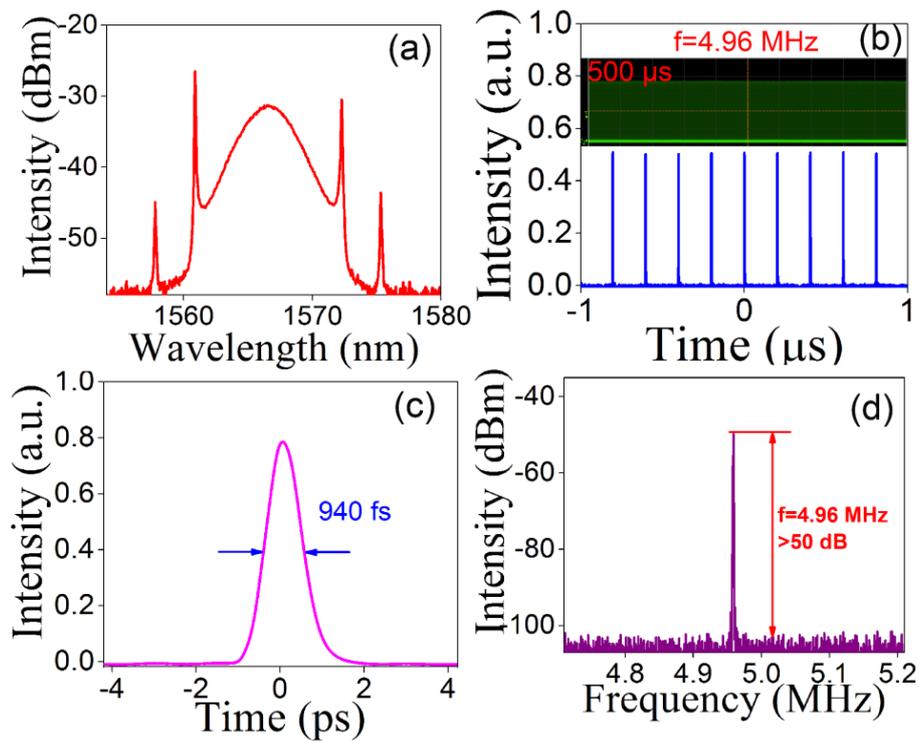

Fig. 7. Mode-locking performance. (a) Spectrum; (b) Pulse-train; (c) Autocorrelation trace; (d) RF spectrum.

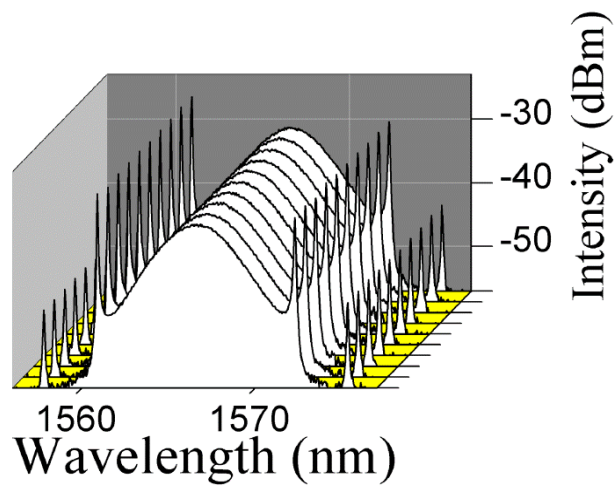

Fig. 8. Repeatedly recorded mode-locked spectrum at a 1-hour interval.

It has been demonstrated that the cavity birefringence could induce the spectral filtering effect in a fiber ring laser, which could be also used to obtain wavelength tunable mode-locking operation by properly adjusting the PCs [37]. Figure 9 depicts the wavelength tuning of mode-locking operation by finely rotating the orientations of the PCs with the fixed pump power of ~70 mW in our fiber laser. The center wavelength of the mode-locked pulse could be continuously tuned from 1532 nm to 1570 nm, covering a wavelength range of 38 nm. The demonstration of the wavelength tuning mode-locking operation indicates that the few-layer BPs could be a broadband saturable absorption 2D material. After the wavelength tunable mode-locking operation, the optical damage of microfiber-based BP SA was tested. To this end, the 980 nm pump power was gradually increased. However, the fiber laser could sustain mode-locking operation well at the maximum pump power of 350 mW in our experiment. In this case, the average output power of mode-locked pulse from the 10% coupling port is 5.6 mW. Therefore, the proposed BPs-deposited microfiber photonic device could at least endure an incident power of 56 mW, indicating that the optical damage of BPs was effectively suppressed by the proposed "lateral interaction scheme". Moreover, we have also checked whether the few-layer BP SA solely contributed to the passive mode locking operation or not. Therefore, the microfiber-based BP SA was purposely removed from the fiber laser. In this case, despite the arbitrary adjustments of PCs and pump power level, no mode-locked pulse could be observed. The comparison results demonstrated that the mode-locking operation of the fiber laser was indeed induced by the saturable absorption effect of BPs.

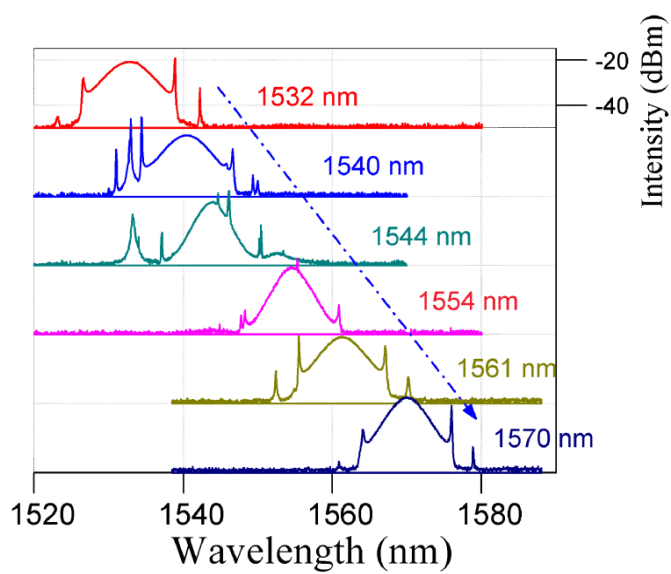

Fig. 9. Wavelength tunable mode-locking operation.

## Conclusion

In conclusion, by depositing few-layer BPs upon the microfiber through the optical trapping approach, we can employ a "lateral interaction scheme" of utilizing the strong optical response of few-layer BPs, through which we could not only strengthen the light-matter interaction thanks to the long interaction distance, but also boost the optical damage threshold of few-layer BP flakes. This BP-deposited microfiber device can withstand strong laser illumination up to ~56 mW. Then by placing the as-fabricated SA device into an erbium-doped fiber laser, we are able to obtain 940 fs mode-locked pulse as well as wideband tunable mode-locking operation. Our results demonstrate that few-layer BPs, as a kind of complentary 2D material for $MoS_2$ and semi-metallic graphene, can further advance the development of photonics based on 2D materials, as a field that investigates the light-matter interaction in 2D materials and the related applications such as light generation, propagation, modulation, and detection.

## Acknowledgments

This work is partially supported by the National Natural Science Foundation of China (Grant Nos. 61435010, 61222505, 61307058, 61378036, 11304101, 11474108, and 11074078), the Key Program of Natural Science Foundation of Guangdong Province, China (Grant No.

2014A030311037), and the Scientific and Technological Innovation Project of Higher Education Institute, Guangdong, China (Grant No. 2013KJCX0051). Z.-C. Luo acknowledges the financial support from the Guangdong Natural Science Funds for Distinguished Young Scholar (Grant No. 2014A030306019), and the Zhujiang New-star Plan of Science & Technology in Guangzhou City (Grant No. 2014J2200008).

## References

1. K. S. Novoselov, D. Jiang, F. Schedin, T. J. Booth, V. V. Khotkevich, S. V. Morozov, and A. K. Geim, "Two-dimensional atomic crystals," PNAS **102**, 10451-10453 (2005).

2. F. Bonaccorso, Z. Sun, T. Hasan, and A. C. Ferrari, "Graphene photonics and optoelectronics," Nat. Photon. **4**, 611-622 (2010).

3. Q. Bao and K. P. Loh, "Graphene photonics, plasmonics, and broadband optoelectronic devices." ACS Nano **6**, 3677-3694 (2012).

4. G. Eda, and S. A. Maier, "Two-Dimensional Crystals: Managing Light for Optoelectronics," ACS Nano **7**(7), 5660-5665 (2013).

5. Q. L. Bao, H. Zhang, Y. Wang, Z. Ni, Y. Yan, Z. Shen, K. P. Loh, and D. Y. Tang, "Atomic-layer graphene as saturable absorber for ultrafast pulsed laser," Adv. Funct. Mater. **19**(19), 3077-3083 (2009).

6. Z. Sun, T. Hasan, F. Torrisi, D. Popa, G. Privitera, F. Wang, F. Bonaccorso, D. M. Basko, and A. C. Ferrari, "Graphene mode-locked ultrafast laser," ACS Nano **4**(2), 803-810 (2010).

7. H. Zhang, Q. Bao, D. Tang, L. Zhao, and K. Loh, "Large energy soliton erbium-doped fiber laser with a graphene-polymer composite mode locker," Appl. Phys. Lett. **95**(14), 141103 (2009).

8. M. Liu, X. B. Yin, E. Ulin-Avila, B. S. Geng, T. Zentgraf, L. Ju, F. Wang, and X. Zhang, "A graphene-based broadband optical modulator," Nature **474**(7349), 64-67 (2011).

9. W. Li, B. G. Chen, C. Meng, W. Fang, Y. Xiao, X. Y. Li, Z. F. Hu, Y. X. Xu, L. M. Tong, H. Q. Wang, W. T. Liu, J. M. Bao, and Y. R. Shen, "Ultrafast all-optical graphene modulator," Nano Lett. **14**(2), 955-959 (2014).

10. F. N. Xia, T. Mueller, Y. M. Lin, A. Valdes-Garcia, and P. Avouris, "Ultrafast graphene photodetector," Nat. Nanotechnol. **4**(12), 839-843 (2009).

11. W. W. Li, X. M. Geng, Y. F. Guo, J. Z. Rong, Y. P. Gong, L. Q. Wu, X. M. Zhang, P. Li, J. B. Xu, G. S. Cheng, M. T. Sun, and L. W. Liu, "Reduced graphene oxide electrically contacted graphene sensor for highly sensitive nitric oxide detection," ACS Nano **5**, 6955-6961 (2011).

12. Q. H. Wang, K. Kalantar-Zadeh, A. Kis, J. N. Coleman, M. S. Strano, "Electronics and optoelectronics of two-dimensional transition metal dichalcogenides," Nat. Nanotechnol. **7**, 699-712 (2012).

13. K. P. Wang, J. Wang, J. T. Fan, M. Lotya, A. O'Neill, D. Fox, Y. Y. Feng, X. Y. Zhang, B. X. Jiang, Q. Z. Zhao, H. Zhang, J. N. Coleman, L. Zhang, and W. J. Blau, "Ultrafast saturable absorption of two-dimensional MoS$_2$ nanosheets," ACS Nano **7**, 9260-9267 (2013).

14. H. Zhang, S. B. Lu, J. Zheng, J. Du, S. C. Wen, D. Y. Tang, and K. P. Loh, "Molybdenum disulfide (MoS$_2$) as a broadband saturable absorber for ultra-fast photonics," Opt. Express **22**, 7249-7260 (2014).

15. S. Wang, H. Yu, H. Zhang, A. Wang, M. Zhao, Y. Chen, L. Mei, and J. Wang, "Broadband Few-Layer MoS$_2$ Saturable Absorbers," Adv. Mater. **26**, 3538-3544 (2014).

16. M. Zhang, R. C. T. Howe, R. I. Woodward, E. J. R. Kelleher, F. Torrisi, G. H. Hu, S. V. Popov, J. R. Taylor, and T. Hasan, "Solution processed MoS$_2$-PVA composite for sub-bandgap mode-locking of a wideband tunable ultrafast er: fiber laser," Nano Research. DOI 10.1007/s12274-014-0637-2 (2014).

17. L. Li, Y. Yu, G. Jun Ye, Q. Ge, X. Ou, H. Wu, D. Feng, X. H. Chen, and Y. Zhang, "Black phosphorus field-effect transistors," Nat. Nanotechnol. **9**, 372-377 (2014).

18. H. Liu, A. T. Neal, Z. Zhu, Z. Luo, X. Xu, D. Tománek, and P. D. Ye, "Phosphorene: An Unexplored 2D Semiconductor with a High Hole Mobility," ACS Nano **8**, 4033-4041 (2014).

19. H. Liu, Y. Du, Y. Deng, and P. D. Ye, "Semiconducting black phosphorus: Synthesis, transport properties and electronic applications," Chem. Soc. Rev. (published online). http://dx.doi.org/10.1039/C4CS00257A (2015).

20. V. Tran, R. Soklaski, Y. Liang, and L. Yang, "Layer-controlled band gap and anisotropic excitons in few-layer black phosphorus," Phys. Rev. B **89**, 235319 (2014).

21. D. Hanlon, C. Backes, E. Doherty, C. S. Cucinotta, N. C. Berner, C. Boland, K. Lee, P. Lynch, Z. Gholamvand, A. Harvey, S. Zhang, K. Wang, G. Moynihan, A. Pokle, Q. M. Ramasse, N. McEvoy, W. J. Blau, J. Wang, S. Sanvito, D. D. Oregan, G. S. Duesberg, V. Nicolosi, and J. N. Coleman, "Liquid exfoliation of solvent-stabilised black phosphorus: applications beyond electronics," http://arxiv.org/abs/1501.01881, accessed 8/1/15.

22. S. B. Lu, L. L. Miao, Z. N. Guo, X. Qi, C. J. Zhao, H. Zhang, S. C. Wen, D. Y. Tang, and D. Y. Fan, "Broadband nonlinear optical response in multi-layer black phosphorus: an emerging infrared and mid-infrared optical material," Opt. Express **23**, 11183-11194 (2015).

23. U. Keller, "Recent developments in compact ultrafast lasers," Nature **424**, 831-838 (2003).

24. Z. Q. Luo, M. Zhou, J. Weng, G. M. Huang, H. Y. Xu, C. C. Ye, and Z. P. Cai, "Graphene-based passively Q-switched dualwavelength erbium-doped fiber laser," Opt. Lett. **35**, 3709-3711 (2010).

25. J. Sotor, G. Sobon, K. Grodecki, and K. M. Abramski, "Sub-130 fs mode-locked Er-doped fiber laser based on topological insulator," Opt. Express **22**, 13244–13249 (2014).

26. K. Wu, X. Zhang, J. Wang, and J. Chen, "463-MHz fundamental mode-locked fiber laser based on few-layer $MoS_2$ saturable absorber," Opt. Lett. **40**, 1374-1377 (2015).

27. Y.-H. Lin, S.-F. Lin, Y.-C. Chi, C.-L. Wu, C.-H. Cheng, W.-H. Tseng, J.-H. He, C.-I Wu, C.-K. Lee, and G.-R. Lin, "Using n- and p-Type $Bi_2Te_3$ Topological Insulator Nanoparticles To Enable Controlled Femtosecond Mode-Locking of Fiber Lasers," ACS Photonics **2**, 481-490 (2015).

28. J. O. Island, G. A. Steele, H. S. J. van der Zant, and A. Castellanos-Gomez, "Environmental instability of few-layer black phosphorus," 2D Materials **1**, 025001 (2014).

29. J. D. Wood, S. A. Wells, D. Jariwala, K. S. Chen, E. K. Cho, V. K. Sangwan, X. Liu, L. J. Lauhon, T. J. Marks, and M. C. Hersam, "Effective Passivation of Exfoliated Black Phosphorus Transistors against Ambient Degradation," Nano Lett. **14**, 6964 (2014).

30. J.-S. Kim, Y. Liu, W. Zhu, S. Kim, D. Wu, L. Tao, A. Dodabalapur, K. Lai, D. Akinwande, "Toward Air-Stable Multilayer Phosphorene Thin-Films and Transistors," http://arxiv.org/abs/1412.0355, accessed 1/12/14.

31. F. Bonaccorso and Z. Sun, "Solution processing of graphene, topological insulators and other 2d crystals for ultrafast photonics," Opt. Mater. Express **4**, 63–78 (2014).

32. H. Liu, A. T. Neal, Z. Zhu, D. Tomanek, and P. D. Ye, "Phosphorene: an unexplored 2D semiconductor with a high hole mobility," ACS NANO **8**, 4033–4041 (2014).

33. Y. Cai, G. Zhang, Y-W. Zhang, "Layer-dependent band alignment and work function of few-layer phosphorene," Sci. Rep. **4**, 6677 (2014).

34. S. B. Lu, C. J. Zhao, Y. H. Zou, S. Q. Chen, Y. Chen, Y. Li, H. Zhang, S. C. Wen, and D. Y. Tang, "Third order nonlinear optical property of $Bi_2Se_3$," Opt. Express **21**, 2072-2082 (2013).

35. Z. C. Luo, M. Liu, H. Liu, X. W. Zheng, A. P. Luo, C. J. Zhao, H. Zhang, S. C. Wen, and W. C. Xu, "2 GHz passively harmonic mode-locked fiber laser by a microfiber-based topological insulator saturable absorber," Opt. Lett. **38**, 5212-5215 (2013).

36. L. E. Nelson, D. J. Jones, K. Tamura, H. A. Haus, and E. P. Ippen, "Ultrashort-pulse fiber ring lasers," Appl. Phys. B **65**, 277–294 (1997).

37. H. Zhang, D. Y. Tang, R. J. Knize, L. M. Zhao, Q. L. Bao, and K. P. Loh, "Graphene mode locked, wavelength tunable, dissipative soliton fiber laser," Appl. Phys. Lett. **96**(11), 111112 (2010).